\documentclass[aps,prl,twocolumn,superscriptaddress,floatfix]{revtex4}
\usepackage{times}
\usepackage{graphicx}
\usepackage{epsfig}
\usepackage{amsmath}
\usepackage{amssymb}
\usepackage[version=3]{mhchem}

\begin{document}

\title{One-dimensional magnetic order in the metal--organic framework Tb(HCOO)$_{\mathbf 3}$}

\author{Daniel R. Harcombe}
\affiliation{Department of Chemistry, University of Oxford, Inorganic Chemistry Laboratory, South Parks Road, Oxford OX1 3QR, U.K.}

\author{Philip G. Welch}
\affiliation{Department of Chemistry, University of Oxford, Inorganic Chemistry Laboratory, South Parks Road, Oxford OX1 3QR, U.K.}
\affiliation{ISIS Facility, Rutherford Appleton Laboratory, Harwell Science and Innovation Campus, Didcot, Oxfordshire OX11 0QX, U.K.}

\author{Pascal Manuel}
\affiliation{ISIS Facility, Rutherford Appleton Laboratory, Harwell Science and Innovation Campus, Didcot, Oxfordshire OX11 0QX, U.K.}

\author{Paul J. Saines}
\email{P.Saines@kent.ac.uk}
\affiliation{School of Physical Sciences, University of Kent, Canterbury CT2 7NH, U.K.}

\author{Andrew L. Goodwin}
\affiliation{Department of Chemistry, University of Oxford, Inorganic Chemistry Laboratory, South Parks Road, Oxford OX1 3QR, U.K.}

\date{\today}
\begin{abstract}
A combination of variable-temperature neutron scattering, reverse Monte Carlo analysis and direct Monte Carlo simulation is used to characterise the emergence of magnetic order in the metal--organic framework (MOF) Tb(HCOO)$_3$ over the temperature range 100\,K to 1.6\,K $=T_{\rm N}$. We show that the magnetic transition at $T_{\rm N}$ involves one-dimensional ferromagnetic ordering to a partially-ordered state related to the triangular Ising antiferromagnet. In this phase, the direction of magnetisation of ferromagnetic chains tends to alternate between neighbouring chains but this alternation is frustrated and is not itself ordered. In neutron scattering measurements this partial order gives rise to Bragg-like peaks, which cannot be interpreted using conventional magnetic crystallography without resort to unphysical spin models. The existence of low-dimensional magnetic order in Tb(HCOO)$_3$ is stabilised by the contrasting strength of inter- and intra-chain magnetic coupling, itself a consequence of the underlying MOF architecture. Our results demonstrate how MOFs may provide an attractive if as yet under-explored platform for the realisation and investigation of low-dimensional physics.
 \end{abstract}


\maketitle




Low-dimensional magnets have long provided an important playground for the discovery and exploitation of unconventional physics \cite{Steiner_1976}---from the earliest studies of soliton excitations in CsNiF$_3$ \cite{Kjems_1978,Mikeska_1991,Ramirez_1982} to contemporary research into quantum information transport in spin-chain compounds \cite{Sahling_2015}. The sensitivity of low-dimensional spin systems to small perturbations results in a rich diversity of phase transitions and complex ordering phenomena. By way of example, the Ising spin-chain compound Ca$_3$Co$_2$O$_6$ exhibits a variety of equilibrium and non-equilibrium states \cite{Aasland_1997,Maignan_2004,Chapon_2009}, characterised by \emph{e.g.}\ long-wavelength incommensurate spin density modulations and field-induced magnetisation plateaux reminiscent of Hofstadter fractalisation \cite{Kageyama_1999,Sebastian_2008}. In the field of low-dimensional magnetism, arguably the strongest scientific interest from both experimental and theoretical perspectives has always been in the limit of strict 1D order \cite{Mekata_1977,Motoyama_1996,Cao_2007,Baumbach_2010,Kudasov_2012,Nandi_2016}. Yet even in canonical systems such as Ca$_3$Co$_2$O$_6$ the divergence of correlation lengths along 1D spin chains is always accompanied by full 3D magnetic order \cite{Chapon_2009,Paddison_2014}. So the realisation and experimental characterisation of genuine partially-ordered low-dimensional spin states remain an important challenge in the field.

It was in this context that we chose to study magnetic order in terbium(III) formate, Tb(HCOO)$_3$. In metal--organic frameworks (MOFs) such as Tb(HCOO)$_3$, magnetically-active transition-metal or rare-earth ions are connected via organic ligands to form extended 3D framework structures. Because organic ligands can support superexchange interactions that span a broad energy range, and because framework design allows controlled incorporation of low-dimensional structural motifs (\emph{e.g.}\ chains, ladders, layers), MOFs are natural candidate hosts for low-dimensional magnetism \cite{Canepa_2013}. Indeed the magnetic response of a number of key MOF families---including formates \cite{Duan_2011}, oxalates \cite{Hursthouse_2004}, and succinates \cite{Saines_2012,Saines_2012b}---can be interpreted in terms of low-dimensional behaviour. However, nearly all of our collective understanding of magnetic order in MOFs is based on indirect experimental techniques, such as magnetic susceptibility, heat capacity, and dielectric constant measurements \cite{Duan_2011,Saines_2012,Saines_2012b,Hursthouse_2004,Huang_2009}. Tb(HCOO)$_3$ is a notable exception: it is one of the few MOFs for which neutron scattering measurements have been used to characterise magnetic structure within both antiferromagnetic ($T<T_{\rm N}\simeq1.6$\,K) and paramagnetic ($T=3$\,K) regimes \cite{Kurbakov_2000,Saines_2015}.


That Tb(HCOO)$_3$ might harbour an unconventional partially-ordered magnetic state at low $T$ is a possibility suggested by the results of these earlier neutron scattering studies. Within the paramagnetic regime, spin correlations are strongly 1D in nature: chains of Tb$^{3+}$ ions couple ferromagnetically, but neighbouring chains interact only weakly \cite{Saines_2015}. On cooling below $T_{\rm N}$, the magnetic Bragg peaks that emerge in the neutron scattering pattern are subtly broader than the non-magnetic Bragg reflections (see SI) and are accompanied by a substantial diffuse scattering component that is highly structured in reciprocal space and so indicative of strongly correlated disorder \cite{Keen_2015}. Moreover, the magnetic structures that were proposed on the basis of conventional crystallographic analysis in both \cite{Kurbakov_2000} and \cite{Saines_2015} require a modulation in Tb$^{3+}$ moment that has no obvious physical origin. In the study of disordered (non-magnetic) materials it is recognised that low-dimensional order often gives Bragg-like scattering features, interpretation of which by conventional crystallographic means leads to spurious structural models \cite{Goodwin_2009,Cairns_2016}.


In this Letter we report a combined variable-temperature neutron scattering, reverse Monte Carlo (RMC), and direct Monte Carlo (DMC) study in which we characterise the emergence of magnetic order in Tb(HCOO)$_3$ on cooling from 100\,K to 1.6\,K. Our key result is that the spin transition at $T_{\rm N}$ involves 1D ferromagnetic (FM) order along Tb$^{3+}$ spin-chains to give a partially-ordered state equivalent to the triangular Ising antiferromagnet (TIA). In this low-$T$ phase, the direction of chain magnetisation tends to alternate between neighbouring chains but this alternation is frustrated and is not itself ordered. This state---which emerges also in DMC simulations---gives rise simultaneously to both Bragg-like and structured diffuse magnetic scattering, with the Bragg intensities indistinguishable from those calculated from the multiple-moment models suggested in previous Rietveld studies. The existence of a well-defined 1D state in Tb(HCOO)$_3$ distinguishes this system from other spin-chain triangular antiferromagnets such as Ca$_3$Co$_2$O$_6$ \cite{Chapon_2009,Paddison_2014}, CsNiF$_3$ \cite{Kamieniarz_1987,Wysin_1986} and CoV$_2$O$_6$ \cite{Nandi_2016} and suggests its low-temperature physics may provide a long-sought-after experimental approximant to ``true'' 1D magnetism. 



Our paper begins with a brief introduction to the crystallography of Tb(HCOO)$_3$. We then present our neutron scattering results, using RMC to quantify the temperature dependence of single-ion anisotropy and pairwise spin correlation functions within the paramagnetic regime. We suggest a simple Hamiltonian that captures the key physics, and use a comparison between DMC and RMC to estimate the magnitude of the exchange and anisotropy terms in this Hamiltonian. We show that the ordering transition at $T_{\rm N}$ in the DMC model is to a partially-ordered (1D) state; moreover this state exactly reproduces the unusual neutron scattering pattern of Tb(HCOO)$_3$ observed at 1.6\,K. We reconcile our interpretation of low-dimensional magnetic order in this phase with the magnetic structure solutions proposed in previous Rietveld analyses. Our paper concludes with a brief comparison of Tb(HCOO)$_3$ with other, more conventional, low-dimensional magnets.

\begin{figure}
\begin{center}
\includegraphics{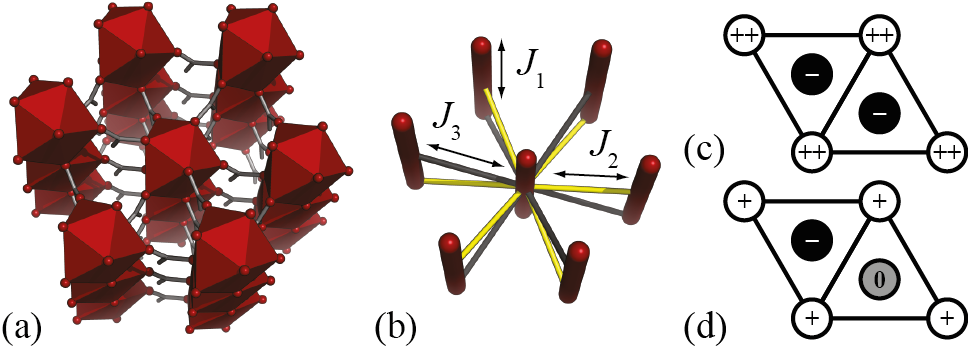}
\end{center}
\caption{\label{fig1} Crystal structure, key magnetic interaction pathways, and candidate magnetic structures of Tb(HCOO)$_3$. (a) Tb$^{3+}$ coordination environments (tricapped trigonal prims; polyhedra) form face-sharing columns parallel to $\mathbf c$. Columns are arranged on a triangular lattice and are connected by formate ions, shown here in stick representation. (b) Intra-chain exchange pathways are ferromagnetic ($J_1$), and the two inequivalent inter-chain exchange pathways $(J_2,J_3$) are collectively antiferromagnetic. Panels (c) and (d) represent the candidate magnetic structures proposed in \cite{Trounov_1997,Saines_2015}. Circles represent Tb chains projected onto the ($a,b$) plane, symbols `$+$' and `$-$' denote ferromagnetic chains with magnetisations along $\mathbf c$ and $-\mathbf c$, respectively, and `0' denotes the absence of any ordered moment.}
\end{figure}

Under ambient conditions, Tb(HCOO)$_3$ crystallises in the rhombohedral space group $R3m$ \cite{Bolotovsky_1990,Lorusso_2013,Saines_2015}. All Tb$^{3+}$ cations are crystallographically equivalent, and are connected by the O atoms of bridging formate ions to form 1D chains that lie coincident with the crystallographic threefold rotation axes [Fig.~\ref{fig1}(a)]. The Tb$\ldots$Tb separation along a given chain is 3.97\,\AA\ with the corresponding Tb--O--Tb angle 105.5$^\circ$---a geometry that favours ferromagnetic coupling and provides a uniaxial crystal field at the Tb$^{3+}$ site (point symmetry $3m$) \cite{Saines_2015}. The chains pack on a perfect triangular lattice with inter-chain separation $a/\sqrt{3}=6.02$\,\AA. Extended Tb--OCO--Tb bridges connect neighbouring chains; one end of the formate bridge has a bifurcated coordination, giving two inequivalent Tb$\ldots$Tb superexchange pathways (distances 6.16 and 6.57\,\AA) for each pair of chain neighbours [Fig.~\ref{fig1}(b)]. There is no experimental evidence for any change in space group symmetry between room temperature and 1.4\,K \cite{Trounov_1997,Saines_2015,Moura_2004}.

Tb(HCOO)$_3$ orders magnetically on cooling to 1.4--1.6\,K \cite{Trounov_1997,Saines_2015}. Conventional symmetry analysis of the magnetic Bragg scattering observed in this ordered phase identifies $P3m^\prime1$ as the unique magnetic space group; the corresponding magnetic cell has the same size as the nuclear cell, but the rhombohedral centering is lost. Two distinct spin ordering patterns (and their linear combinations) are consistent with the observed magnetic Bragg reflection conditions and intensities. Both cases demand single-ion anisotropy with local moments aligned parallel to the $c$-axis, both require FM correlations along the Tb chains, and both correspond to antiferromagnetic (AFM) ordering patterns with zero net magnetisation. In one model, one third of the chains has twice the ordered moment of the other two thirds; the magnetisation direction of these two components oppose [Fig.~\ref{fig1}(c)]. In the second model, one third of the chains has no ordered moment, and the other two thirds alternate their magnetisation directions [Fig.~\ref{fig1}(d)]. We will come to show that these models are an artefact of applying conventional crystallographic approaches to a state that is not 3D ordered, and that neither model describes the true magnetic structure below $T_{\rm N}$.

We study the emergence of magnetic order in Tb(DCOO)$_3$, making use of variable-temperature neutron total scattering data measured by the high-resolution WISH diffractometer at ISIS \cite{Chapon_2011}. The same powdered sample used in \cite{Saines_2015} was placed into vanadium cans of 6\,mm diameter and loaded on the diffractometer. The sample temperature was varied between 300 and 1.6\,K using a low-background $^4$He cryostat. Typical data collection strategies involved neutron counts of \emph{ca} 20\,$\mu$A\,h. Data were corrected using standard protocols as implemented in the Mantid software, merging banks at constant scattering geometry to improve counting statistics. The scattering from an empty 6\,mm vanadium can was also measured under identical conditions, processed using the same protocols, and used for background subtraction. The nuclear scattering contribution is well confined to nuclear Bragg peaks, which could be fitted using GSAS \cite{vonDreele_1986,Larson_2001} and subtracted from the total scattering data collected at the 27.1$^\circ$ and 58.3$^\circ$ banks. These corrected data were merged, placed on an absolute scale (using the GSAS scale factor), and rebinned at intervals of $\Delta Q=0.02$\,\AA$^{-1}$. The total usable $Q$-range spanned 0.1--3.7\,\AA, although within this range a total of five regions contaminated with nuclear scattering contributions were excised and omitted from our subsequent analysis.

\begin{figure}
\begin{center}
\includegraphics{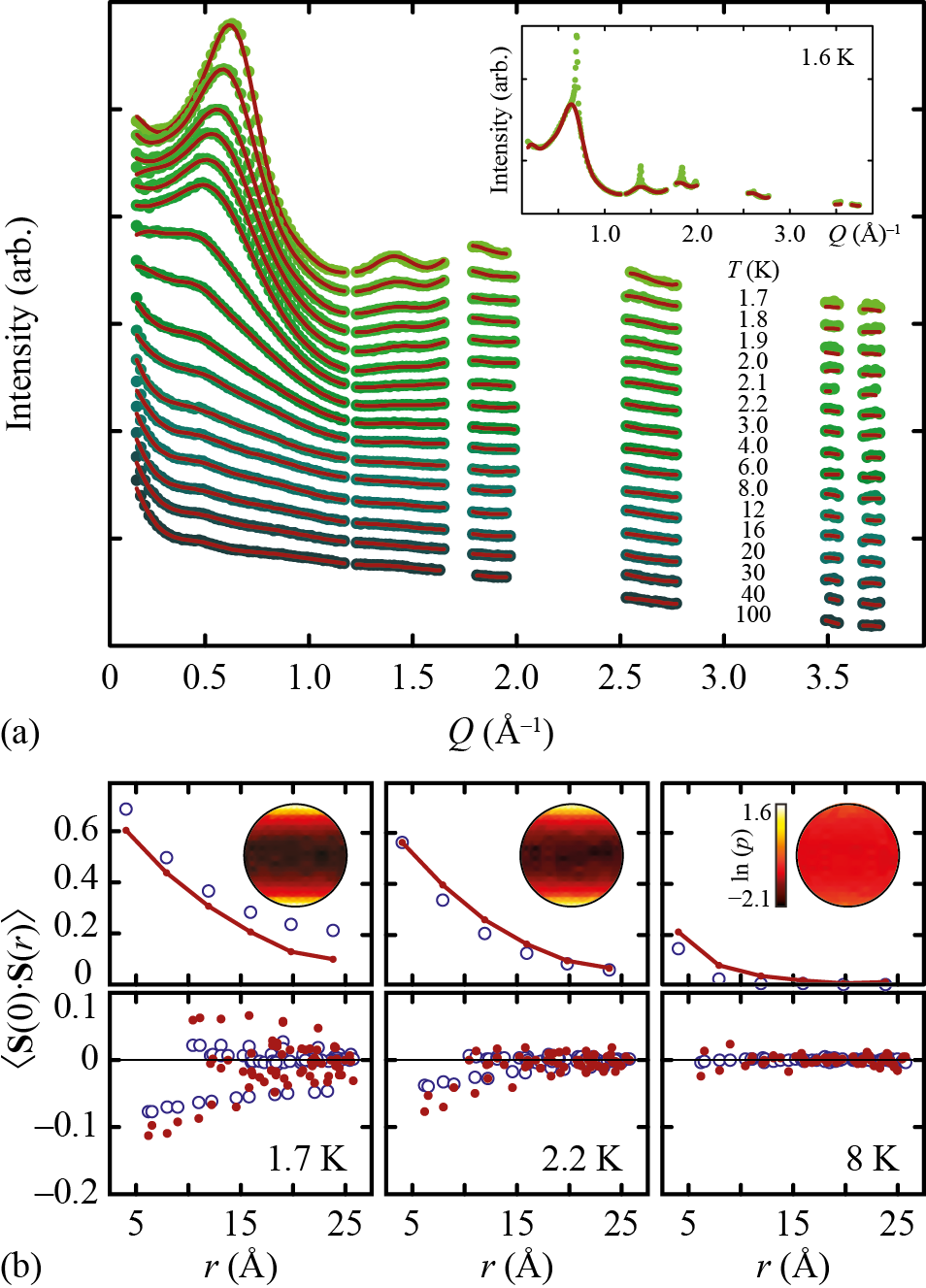}
\end{center}
\caption{\label{fig2} Temperature-dependent magnetic scattering in Tb(DCOO)$_3$ and its SPINVERT analysis. (a) Scattering data are shown as filled circles, with SPINVERT fits shown as red solid lines. Regions of the scattering pattern contaminated with nuclear scattering have been excluded. The inset shows scattering data collected at 1.6\,K within the partially ordered regime together with a small-box SPINVERT fit. (b) Single-ion anisotropy and pairwise correlation functions at three representative temperatures spanning the paramagnetic regime studied here. The single-ion spin orientation functions are shown in stereographic projection, with colours indicating relative distribution probability $p(\theta,\phi)=\rho(\theta,\phi)/{\rm d}(\cos\theta){\rm d}\phi$; here $\rho(\theta,\phi)$ is the fraction of total spins within the angular region ${\rm d}(\cos\theta),{\rm d}\phi$. Spin correlation functions have been separated into intra-chain (top panels) and inter-chain (bottom panels) terms. RMC and DMC results are shown using filled red and open blue circles, respectively; uncertainties are smaller than the symbols.}
\end{figure}

Our corrected data are shown in Fig.~\ref{fig2}(a). We find no evidence of magnetic order for any temperature greater than 1.6\,K, and obtain excellent consistency with the earlier studies of Refs.~\citenum{Trounov_1997,Saines_2015}. We proceeded to fit these data using the SPINVERT implementation of magnetic RMC \cite{Paddison_2012,Paddison_2013}, making use of the analytical approximation to the Tb$^{3+}$ form factor given in Ref.~\citenum{Brown_1995}. For our initial refinements, we used spin configurations that represented a supercell of the nuclear $R3m$ cell given by the transformation
\begin{equation}
\left[\begin{array}{l}\mathbf a\\ \mathbf b\\ \mathbf c\end{array}\right]_{\textrm{RMC}}=\left[\begin{array}{ccc}3&-3&0\\ 5&5&0\\ 0&0&13\end{array}\right]\left[\begin{array}{l}\mathbf a\\ \mathbf b\\ \mathbf c\end{array}\right]_{R3m}.
\end{equation}
This choice of supercell yields orthogonal and approximately isotropic configurations of size $\sim$(50\,\AA$)^3$ containing 1170 spins each; larger configurations gave equivalent results but at greater computational expense. The SPINVERT fits obtained are shown in Fig.~\ref{fig2}(a). We find the data are well modelled at all temperatures within the paramagnetic regime, with a marginal improvement if we use spins with Heisenberg rather than Ising degrees of freedom (see SI).  A SPINVERT fit to the 1.6\,K data set---\emph{i.e.} within the ordered regime---accounts satisfactorily for the diffuse component but cannot reproduce the Bragg scattering [see inset to Fig.~\ref{fig2}(a)]. This is to be expected \cite{Paddison_2013} as the reciprocal-space resolution $\Delta Q\sim0.01$\,\AA$^{-1}$ required to describe Bragg features is many times smaller than the resolution afforded by our RMC cells ($\Delta Q\simeq2\pi/r_{\textrm{max}}=0.2$\,\AA$^{-1}$). We will return to this point later in our study.


The temperature dependence of spin orientations and pairwise spin correlations is illustrated in Fig.~\ref{fig2}(b) for three representative temperatures (1.7, 2.2, and 8\,K). The Ising-like anisotropy noted in \cite{Saines_2015} is clearly evident at both 1.7 and 2.2\,K, but is not obvious at 8\,K. These RMC results represent a lower bound on the true single-ion properties as RMC necessarily underestimates anisotropy \cite{Paddison_2013,Paddison_2015}; indeed we cannot rule out that Ising-like anisotropy persists to much higher temperatures since RMC fits using Ising spins also provide acceptable fits to the neutron scattering data (see SI). Irrespective of the degree of anisotropy what certainly evolves at low temperatures is the extent of FM correlations along Tb chains. At the very lowest temperatures, AFM inter-chain correlations also become significant, as evidenced by the negative values of $\langle\mathbf S(0)\cdot\mathbf S(r)\rangle$ for $r=6.16$ and 6.57\,\AA. These interactions are necessarily frustrated, and lead to net FM correlations for next-nearest chain neighbours (\emph{e.g.}, $r=a=10.42$\,\AA).

On the basis of these correlation functions, we anticipated that the basic spin physics of our system might be captured by a combination of Ising-like single-ion anisotropy, FM intra-chain interactions and AFM inter-chain interactions. The simplest Hamiltonian containing these ingredients is
\begin{equation}\label{ham}
\mathcal H=J_\parallel\sum_{\langle i,j\rangle}\mathbf S_i\cdot\mathbf S_j+J_\perp\sum_{\langle\langle i,j\rangle\rangle}\mathbf S_i\cdot\mathbf S_j-D\sum_iS_{iz}^2,
\end{equation}
where $\langle\cdots\rangle$ and $\langle\langle\cdots\rangle\rangle$ denote sums over neighbouring spins within and between chains, respectively, $J_\parallel=J_1<0$ and $J_\perp=J_2\equiv J_3>0$ are as shown in Fig.~\ref{fig1}(b), and $D>0$ is the single-ion term. To the best of our knowledge, there is no general theory for this interaction model on the rhombohedral lattice that would allow us to extract the $J_i,D$ parameters directly from the experimental spin correlation functions, even if specific realisations are well understood \cite{Loveluck_1975,Sasaki_1982}. Consequently, our approach is to use Eq.~\eqref{ham} to drive DMC simulations with different parameter sets to attempt to reproduce the basic temperature dependence of the experimental spin correlation functions, and so identify a set of $J_i,D$ values that capture the key behaviour of paramagnetic Tb(HCOO)$_3$.

Using a grid search approach we tested a variety of candidate $J_i,D$ values, carrying out DMC simulations using the code developed in \cite{Paddison_2015} and ranking parameter sets according to the fidelity of reproduction of the pairwise spin correlation functions at all measured temperatures. Our DMC spin configurations were the same size and geometry as in the original SPINVERT refinements. Simulations were initialised with random spin orientations, with $T$ systematically lowered from 100\,K to 1.7\,K following equilibration at each step. Simulations were repeated 100 times and the correlation functions averaged over these independent runs. We found the closest match to experiment for $J_\parallel=1.5(5)$\,K, $J_\perp=-0.03(1)$\,K, $D=70(20)$\,K [Fig.~\ref{fig2}(b)] (note that we have subsumed the $S^2$ term within these $J_i,D$ values). The match to AFM inter-chain interactions would likely be improved by distinguishing $J_2$ and $J_3$, but we have not needed this additional complexity for the purposes of this study. We do note, however, that the $J_\parallel$ and $D$ parameters showed strong covariance, such that reasonable fits could be obtained with larger $J_\parallel$ and smaller $D$. Because the RMC spin orientation distributions underestimate anisotropy, we cannot use these distributions to help quantify $D$ (other than to act as a flag were the value too low, which is not the case here). Nevertheless the value of $D$ we obtain is consistent with that obtained elsewhere for Tb$^{3+}$ in axial crystal fields \cite{Gingras_2000}, and the qualitative behaviour of the DMC simulations themselves is the same for all acceptable $J_\parallel,D$ combinations we identified (see SI).

\begin{figure}
\begin{center}
\includegraphics{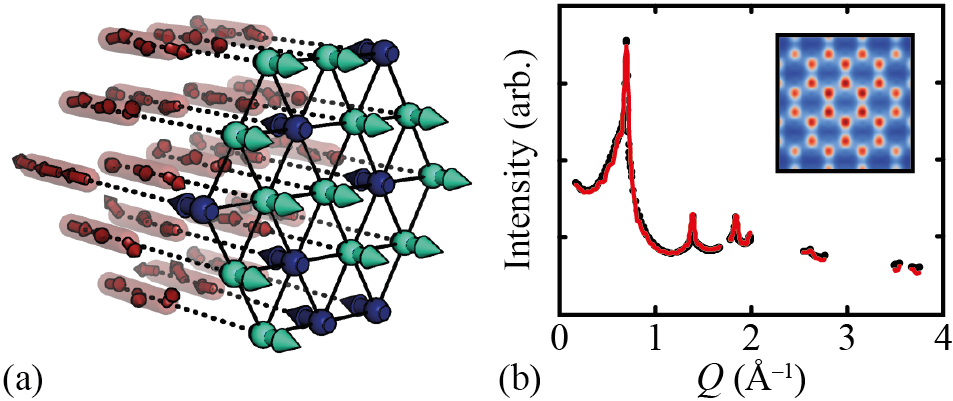}
\end{center}
\caption{\label{fig3} 1D magnetic order in Tb(HCOO)$_3$. (a) A fragment of the DMC spin configurations at a simulation temperature of 1.5\,K is represented in red: spins within a given chain share a common magnetisation direction parallel to the chain axis. Projection of these magnetisation directions onto the underlying triangular lattice on which chains are arranged gives a realisation of the triangular Ising antiferromagnet (green and blue spins). (b) Comparison of experimental neutron scattering data (black symbols) with the neutron scattering pattern calculated for a TIA state as shown in (a) (red symbols). Bragg-like features appear at the regions of reciprocal space associated with maxima in the TIA diffuse scattering pattern (inset).}
\end{figure}

As in the real system itself, this particular Hamiltonian exhibits an ordering transition on cooling below $\sim$1.6\,K for the various $J_i,D$ values consistent with our neutron scattering measurements of the paramagnetic phase. Direct interrogation of the relevant DMC configurations reveals the low-$T$ state to contain only partial order [Fig.~\ref{fig3}(a)]. Individual chains exhibit FM order with their magnetisation aligned either parallel or antiparallel to $\mathbf c$; the chain magnetisation tends to alternate between neighbouring chains but this alternation is frustrated by the underlying triangular lattice and so there is no 3D magnetic order. In other words, the state is a realisation of the TIA where individual Ising variables correspond to collective chain magnetisations. Extending this model to substantially larger configurations (\emph{ca} (20\,nm)$^3$, or $\sim$10$^5$ spins) allows calculation of a corresponding neutron scattering pattern of sufficient reciprocal-space resolution to account at once for both Bragg and diffuse scattering. Crucially, we find that the intermediate order state identified by DMC simulations can account for the entire experimental scattering pattern, including the appearance of Bragg-like peaks with the correct reflection conditions and relative intensities [Fig.~\ref{fig3}(b)]. By construction, this model requires no unphysical modulation of the Tb$^{3+}$ moment. Rather the \emph{magnitude} of chain magnetisation is homogeneous throughout the configuration; our model contains an average spin projection $|\langle S_{z}\rangle|=0.662(4)$ along the chain axes with standard deviation = 0.136.


So our analysis suggests that the magnetically ordered state of Tb(HCOO)$_3$ accessed below 1.6\,K is 1D in nature, such that the remnant spin degrees of freedom map this system onto the 2D TIA. The sharp reflections observed in the neutron scattering patterns are not true Bragg reflections, but correspond to maxima in the continuous scattering function [inset to Fig.~\ref{fig3}(b)] with precisely the form expected for TIA phases \cite{Wojtas_2009}. It was shown in \cite{Yoon_2014} that interpretation of these ``reflections'' using conventional crystallographic approaches leads to a three-sublattice average structure model identical to that portrayed in Fig.~\ref{fig1}(c). Consequently, the modulation in magnitude of ordered moment from chain to chain given by that model is an artefact of applying conventional crystallographic tools to a state with no 3D magnetic order. We note that the magnetic scattering seen here bears strong analogy to the X-ray diffraction patterns of fibre assemblies (\emph{e.g.}\ dense carbon nanotube samples), which also show 1D order \cite{Reznik_1995}.


Our identification of 1D magnetic order in Tb(HCOO)$_3$ has a number of implications. For example, we anticipate by analogy to CsNiF$_3$ the signature of emergent phenomena in its excitation spectrum \cite{Kjems_1978,Mikeska_1991,Ramirez_1982}; indeed the contrast in single-ion anisotropy between these two systems (easy-axis \emph{vs} easy-plane) provides a means by which to test key aspects of the underlying theory of anisotropic 1D ferromagnets \cite{Steiner_1976,Loveluck_1975}. Although the much-studied Ca$_3$Co$_2$O$_6$ is conceptually similar to Tb(HCOO)$_3$, the key distinction between the two is the order-of-magnitude difference in $J_\parallel/J_\perp$ values that stabilises the intermediate-order state in the latter \cite{Paddison_2014}. Nevertheless, as is the case for Ca$_3$Co$_2$O$_6$, Tb(HCOO)$_3$ is likely to show anomalous response to applied magnetic field; indeed magnetisation plateaux may explain its high performance as a magnetocaloric material \cite{Saines_2015,Nandi_2016}. From a materials design perspective, the 1D behaviour of Tb(HCOO)$_3$ is a direct consequence of the underlying MOF architecture. Chemical substitution of formate for longer bridging ligands may allow even more extreme $J_\parallel/J_\perp$ values to be realised, presumably stabilising 1D behaviour over a larger temperature range. There is scope too for substitution at the Tb$^{3+}$ site, since the lanthanide formates are isostructural \cite{Bolotovsky_1990}. Because Gd(HCOO)$_3$ is known to show AFM coupling \emph{within} chains, the solid solution Gd$_{1-x}$Tb$_x$(HCOO)$_3$ \cite{Saines_2015} may provide an attractive entry point for the realisation of random-chain 1D magnets \cite{Nguyen_1996}.


\begin{acknowledgments}
The authors gratefully acknowledge financial support from the Glasstone Bequest (Oxford), the E.R.C. (Grant 279705), E.P.S.R.C. and S.T.F.C. This work was made possible by access to the WISH diffractometer at ISIS. We thank S. Bovill and L. Timm for assistance with the neutron scattering experiments, J. A. M. Paddison (Georgia Tech.) for illuminating discussions and for the use of his DMC code.
\end{acknowledgments}



\end{document}